%% file: main.tex
\documentclass[conference]{IEEEtran}
\IEEEoverridecommandlockouts

\def\BibTeX{{\rm B\kern-.05em{\sc i\kern-.025em b}\kern-.08em
    T\kern-.1667em\lower.7ex\hbox{E}\kern-.125emX}}

\input{custom_style}

\usepackage{array} 
\usepackage{graphicx}
\usepackage[symbol]{footmisc}
\usepackage{svg}
\usepackage{adjustbox}
\usepackage{multirow}

\usepackage{algorithm}
\usepackage{algpseudocode}
\usepackage{physics}

\usepackage[T1]{fontenc}    
\usepackage[utf8]{inputenc} 
\usepackage{hyperref}       
\usepackage{url}            
\usepackage{booktabs}       
\usepackage{amsfonts}       
\usepackage{nicefrac}       
\usepackage{microtype}      
\usepackage{xcolor}         

\usepackage{amsmath,amsfonts,amssymb,amsthm,mathtools,nicefrac,math}

\usepackage{listings} 
\usepackage{xcolor}   

\usepackage[margin=1in]{geometry}
\usepackage{qcircuit}

\usepackage{booktabs}

\begin{document}

\title{Quantum parallel information exchange (QPIE) hybrid network with transfer learning\\
}

\makeatletter
\newcommand{\linebreakand}{%
  \end{@IEEEauthorhalign}
  \hfill\mbox{}\par
  \mbox{}\hfill\begin{@IEEEauthorhalign}
}
\makeatother
\author{\IEEEauthorblockN{Ziqing Guo}
\textit{Texas Tech University}\\
Lubbock, USA \\
ziqguo@ttu.edu
\and
\IEEEauthorblockN{Alex Khan}
\textit{National Quantum Laboratory}\\
Maryland, USA \\
askhan@umd.edu
\and
\IEEEauthorblockN{Victor S. Sheng}
\textit{Texas Tech University}\\
Lubbock, USA \\
victor.sheng@ttu.edu
\linebreakand
\IEEEauthorblockN{Shabnam Jabeen}
\textit{University of Maryland}\\
Maryland, USA \\
jabeen@umd.edu
\and
\IEEEauthorblockN{Ziwen Pan}
\textit{Texas Tech University}\\
Lubbock, USA \\
ziwen.pan@ttu.edu
}

\maketitle

\input{sections/abstract_v3}
\input{sections/01_introduction}

\input{sections/02_results}
\input{sections/04_methods}
\input{sections/06_related_work}
\input{sections/03_discussion}

\input{sections/acknowledgements}

\bibliographystyle{IEEEtran}
\bibliography{bibs/main, bibs/the, bibs/tools, bibs/qml}
\end{document}

%% file: custom_style.tex
\newcommand{\qp}{\textbf{\texttt{QPIE}}}

\newcommand{\resnet}{\textbf{\texttt{ResNet-18}}}
\newcommand{\resnets}{\textbf{\texttt{ResNet-50}}}
\newcommand{\sv}{\textbf{\texttt{SV1}}}
\newcommand{\aria}{\textbf{\texttt{Aria1}}}
\newcommand{\pen}{\textbf{\texttt{Pennylane}}}
\newcommand{\cq}{\textbf{\texttt{CUDA-Q}}}
\newcommand{\amazon}{\textbf{\texttt{Amazon}}}
\newcommand{\ionq}{\textbf{\texttt{IonQ}}}
\usepackage{array} 
\usepackage{graphicx}  
\usepackage[symbol]{footmisc}
\usepackage{svg}
\usepackage{adjustbox}
\usepackage{multirow}
\usepackage{subcaption}
\usepackage{algorithm}
\usepackage{algpseudocode}
\usepackage{physics}

\usepackage[T1]{fontenc}    
\usepackage[utf8]{inputenc} 
\usepackage{hyperref}       
\usepackage{url}            
\usepackage{booktabs}       
\usepackage{amsfonts}       
\usepackage{nicefrac}       
\usepackage{microtype}      
\usepackage{xcolor}         

\usepackage{amsmath,amsfonts,amssymb,amsthm,mathtools,nicefrac,math}

\usepackage{listings} 
\usepackage{xcolor}   

\usepackage[margin=1in]{geometry}
\usepackage{environ}
\newcommand{\myqctmp}[2][0.25]{\Qcircuit @C=#2em @R=#1em @!R}
\NewEnviron{myqcircuit}[1][0.25]{\vcenter{\myqctmp[#1]{0.7} {\BODY}}} 
\NewEnviron{myqcircuit*}[2]{\vcenter{\myqctmp[#1]{#2} {\BODY}}}

\usepackage{booktabs}

\usepackage{cite}               
\usepackage{breakurl}           
\usepackage{url}                
\usepackage{xcolor}             
\usepackage[]{hyperref}         
\hypersetup{                    
  colorlinks,
  linkcolor={green!80!black},
  citecolor={red!70!black},
  urlcolor={blue!70!black}
}
\usepackage[capitalize,nameinlink]{cleveref}

\usepackage{comment}

\usepackage{makecell}
\newcommand{\thickhline}{\Xhline{2.5\arrayrulewidth}}

%% file: sections/abstract_v3.tex
\begin{abstract}
Quantum machine learning (QML) has emerged as an innovative framework with the potential to uncover complex patterns by leveraging quantum systems ability to simulate and exploit high-dimensional latent spaces, particularly in learning tasks.
Quantum neural network (QNN) frameworks are inherently sensitive to the precision of gradient calculations and the computational limitations of current quantum hardware as unitary rotations introduce overhead from complex number computations, and the quantum gate operation speed remains a bottleneck for practical implementations.
In this study, we introduce quantum parallel information exchange (\qp) hybrid network, a new non-sequential hybrid classical quantum model architecture, leveraging quantum transfer learning by feeding pre-trained parameters from classical neural networks into quantum circuits, which enables efficient pattern recognition and temporal series data prediction by utilizing non-clifford parameterized quantum gates thereby enhancing both learning efficiency and representational capacity.
Additionally, we develop a dynamic gradient selection method that applies the parameter shift rule on quantum processing units (QPUs) and adjoint differentiation on GPUs. 
Our results demonstrate model performance exhibiting higher accuracy in ad-hoc benchmarks, lowering approximately 88\% convergence rate for extra stochasticity time-series data within 100-steps, and showcasing a more unbaised eigenvalue spectrum of the fisher information matrix on CPU/GPU and IonQ QPU simulators. 
\end{abstract}

\begin{IEEEkeywords}
quantum machine learning, parameterized quantum circuit, quantum embedding, hybrid quantum model
\end{IEEEkeywords}

%% file: sections/01_introduction.tex
\section{Introduction}
Quantum machine learning (QML) has the potential to improve the efficiency of training and evaluation, leveraging quantum circuit simulation (QCS) to directly model the mathematical structure of complex quantum states \cite{b1}.
The rapid evolution of advanced hardware resources \cite{alvarez2018quantum}, combined with advancements in quantum error correction techniques \cite{b5}, has empowered quantum-enhanced computing to tackle a wide range of tasks, such as optimization \cite{abbas2024challenges}, image recognition \cite{Balewski2024quantum_parallel}, time-series forecasting \cite{chen2020temporal}, and complex data classification \cite{qnn1}, with improved reliability and scalability.
In the era of NISQ \cite{b34}, such variational quantum circuits (VQCs) \cite{chen2020quantumconvolutionalneuralnetworks} demonstrate their capability in learning tasks because quantum circuits provide a highly expressive framework for encoding problem-specific parameters using trainable quantum gates (unitary gates).
This results in efficient exploration of the solution space, leveraging quantum superposition and entanglement, similar to the behavior of quantum neural networks (QNNs) \cite{qnn2, mari2020transfer}.

\begin{figure}[htbp]
\centering
\includegraphics[scale=0.85]{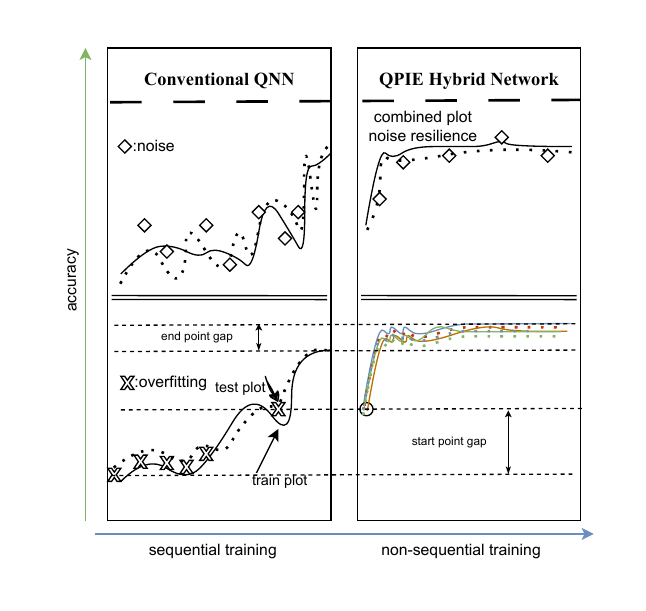}
\caption{\label{fig:story} Unavoidable quantum noise from QPUs, inconsistent training schemes, and prolonged convergence periods currently limit the common QNN scheme. In contrast, QPIE enhances initial training accuracy through parallel learning predictions, thereby decreasing the model sensitivity to quantum noise.}
\end{figure}

Despite the potential of QML for nonlinear, high-dimensional temporal tasks, its practical implementation faces challenges such as the deeper circuit calculation needing for dynamic anstaz management \cite{patil2022variational}, increasing demand for quantum hardware requirements \cite{b21}, gradient vanishing \cite{wang2021noise}, and vulnerability to decoherence and gate control errors in particular trapped-ion based QPUs \cite{bruzewicz2019trapped}.
For instance, the sequential layered architecture of QNNs can amplify gradient vanishing, particularly in deep circuits with over 28 qubits and more than 1,000 entangling gates, rendering quantum optimization inefficiency \cite{s2}. Consequently, there is an increasing demand for tailored design patterns that enable efficient training while achieving higher accuracy for the learning applications.

\begin{table*}[htbp]
\centering
\caption{Benchmark on classification tasks and average training time (CPU/GPU) for local simulators. We test Hymenoptera, Brain Tumor, and MNIST dataset with additional \amazon\ \sv\ and \ionq\ \aria\ quantum simulators (Results labeled with NN* indicate we use classical models without quantum layers.)}
\label{tab:results}
\adjustbox{max width=\linewidth}{
\setlength{\tabcolsep}{8pt} 
\def\arraystretch{1.25}
\begin{tabular}{ccccccccccccc}
\thickhline
\multirow{2}{*}{\textbf{Model}}  & 
\multicolumn{11}{c}{\textbf{Accuracy ($\uparrow$)}}  \\ \cline{2-13} 
&& {\textbf{Moon}} & {\textbf{Spiral}} & & {\textbf{Hymenoptera}} & & & {\textbf{Brain Tumor}} & && {\textbf{MNIST}} \\ 
\midrule
NN* && 0.93 & 0.90 & & 0.87 & & & 0.76 & & &0.81 \\ 
\midrule
& \begin{tabular}[c]{@{}c@{}}\textbf{Pre-trained}\\ \textbf{Model}\end{tabular} & \begin{tabular}[c]{@{}c@{}}Local\\ 28 qubits\end{tabular} & \begin{tabular}[c]{@{}c@{}}Local\\ 28 qubits\end{tabular} & \begin{tabular}[c]{@{}c@{}}Local\\ 28 qubits\end{tabular} & \begin{tabular}[c]{@{}c@{}}SV1\\ 34 qubits\end{tabular} & \begin{tabular}[c]{@{}c@{}}Aria1\\ 25 qubits\end{tabular} & \begin{tabular}[c]{@{}c@{}}Local\\ 28 qubits\end{tabular} & \begin{tabular}[c]{@{}c@{}}SV1\\ 34 qubits\end{tabular} & \begin{tabular}[c]{@{}c@{}}Aria1\\ 25 qubits\end{tabular}& \begin{tabular}[c]{@{}c@{}}Local\\ 28 qubits\end{tabular} & \begin{tabular}[c]{@{}c@{}}SV1\\ 34 qubits\end{tabular} & \begin{tabular}[c]{@{}c@{}}Aria1\\ 25 qubits\end{tabular} \\ 
\cline{2-13} 
QNN & Resnet-18 & 0.94 & 0.91 & 0.88 & 0.86 & 0.84 & 0.81 & 0.78 & 0.79 & 0.91 & 0.84 & 0.83\\ 
& & 8.2s & 16.1s & 18.1s & 32.7s & 46.9s & 21.4s & 46.4s & 48.9s & 40.4s & 48.6s & 40.7s\\ 
QNN & Resnet-50 & 0.95 & 0.92 & 0.91 & 0.88 & 0.87 & 0.84 & 0.81 & 0.80 & 0.92 & 0.88 & 0.89 \\ 
& & 12.8s & 25.6s & 28.8s & 51.2s & 73.6s & 33.6s & 73.6s & 76.8s & 64s & 76.8s & 64s\\ 
\textbf{\qp\ } & Resnet-18 & \textbf{0.95/0.99} & \textbf{0.96/0.99} & \textbf{0.92/0.95} & \textbf{0.93} & \textbf{0.92} & \textbf{0.87/0.91} & \textbf{0.86} & \textbf{0.88} & \textbf{0.93/0.95} & \textbf{0.88} & \textbf{0.83}\\ 
& & 9.0s/5.7s & 18.5s/11.3s & 20.8s/12.7s & 22.9s & 32.8s & 23.3s/15.0s & 32.5s & 34.2s & 44.2s/28.3s & 34.0s & 28.5s \\ 
\textbf{\qp\ } & Resnet-50 & \textbf{0.97/0.99} & \textbf{0.96/0.99} & \textbf{0.94/0.98} & \textbf{0.93} & \textbf{0.93} & \textbf{0.88/0.93} & \textbf{0.88} & \textbf{0.90} & \textbf{0.93/0.97} & \textbf{0.89} & \textbf{0.96}\\ 
& & 14.2s/9.0s & 28.4s/18.0s & 32.7s/18.0s & 35.2s & 42.6s & 36.9/20.2s & 42.6s & 47.0s & 72.8s/44.8s & 47.0s & 44.8s\\ 
\thickhline
\end{tabular}%
}
\end{table*}

To address these challenges, this paper presents the \qp\ hybrid network, which integrates pre-trained model weights and dynamic data re-uploading mid-measurement circuits. Our approach enables multi-GPU acceleration with dynamic gradient calculation support, effectively solving nonlinear temporal and prediction tasks. This architecture supports parallel execution, demonstrates rapid convergence during training, and is robust to noise, enabling efficient and reliable model performance, as illustrated in \cref{fig:story}.

%% file: sections/02_results.tex
\section{Results}
\label{sec:results}
In this section, we describe the \qp\ framework versatility test across three tasks and demonstrate the capability of our model. We present the results in three separate categories:
1) We first assess the model performance utilizing different datasets, including non-linear spiral and moon benchmarks, brain tumors, ants and bees, and MNIST \cite{cameron1903hymenoptera, cohen2017emnist, cong2022chip, abiwinanda2019brain}. We note that the computational power of current quantum neural networks (QNNs) depends on the number of quantum operations. Therefore, we select datasets that are maximally compatible with limited quantum circuit depth.
2) Next, we test the temporal prediction using nonlinear autoregressive moving average (NARMA5) and (NARMA10) \cite{appeltant2011information} to evaluate the \qp\ gradients convergence efficiency.
3) To further evaluate the model's capacity to explore the Hilbert space, we employ the Fisher Information Matrix (FIM)~\cite{meyer2021fisher} to assess the sensitivity of the model within its latent space.

\input{sections/result_01_image}
\input{sections/result_02_series}
\input{sections/result_03_model_capability}

%% file: sections/result_01_image.tex
\subsection*{Pattern recognition performance}
\label{res1}
We demonstrate the \qp\ setup with pre-trained non-sequential neural network \resnet\ \cite{resnet} by testing on the local GPU-accelerated quantum simulator, cloud on-demand \amazon\ \sv\ simulator\footnote{https://aws.amazon.com/braket/quantum-computing-research/}, and \aria\ QPU \cite{ionq} in \cref{tab:results}.
We observe the \qp\ results outperform the conventional NN (see \cref{med:exp} for classic neural network settings) and sequential QNN in terms of running time and model accuracy after training for 100 epochs (See \cref{med:hcqm} for model initialization).

In essence, we first divide our test into two categories: binary non-linear pattern recognition (BR) and multi-label classification (MC). Our results demonstrate that \qp\ is not hindered by the difference between BR and MC because the dynamic measurement gates in the VQC of \qp\ provide more capacity for latent space pattern recognition. Consequently, the time consumption compared to classical QNN is approximately 10\% longer, as shown in \cref{tab:results}. 
To reduce the time consumption, we test the model utilizing GPU-based simulators, achieving up to a 30-fold speed-up without losing accuracy. Conversely, we present the results utilizing QPU with comparable time consumption but lower accuracy due to the coherent noise and gate fidelity limitations of trapped ion quantum simulators \cite{ionq}. Notably, the GPU simulators (right) exhibit higher accuracy than the CPU (left) because we introduce density matrix depolarization noise.

We note that current transformer-based large language models are trendily used in generation tasks \cite{croitoru2023diffusion, huang2024gan}; however, in the QML field, due to the limitations of qubits, the VQC cannot inherit a large number of parameters. Inspired by QTL \cite{mari2020transfer}, we design two sets of architectures by selecting \resnet\ and \resnets\ with pre-trained ImageNet parameters \cite{stevens2020deep}, which interconnect with our VQC defined in \qp\ (see \cref{med:qpie}).
We observe that \qp\ with \resnet\ achieves a trade-off between accuracy and time consumption, as \resnets\ introduces a larger overhead in our experiments, despite having over 40.0 GB of memory within a single GPU computation node. Furthermore, an overfitting issue frequently arises when using \resnets.

\begin{figure}[htbp]
    \centering
    \includegraphics[width=\linewidth]{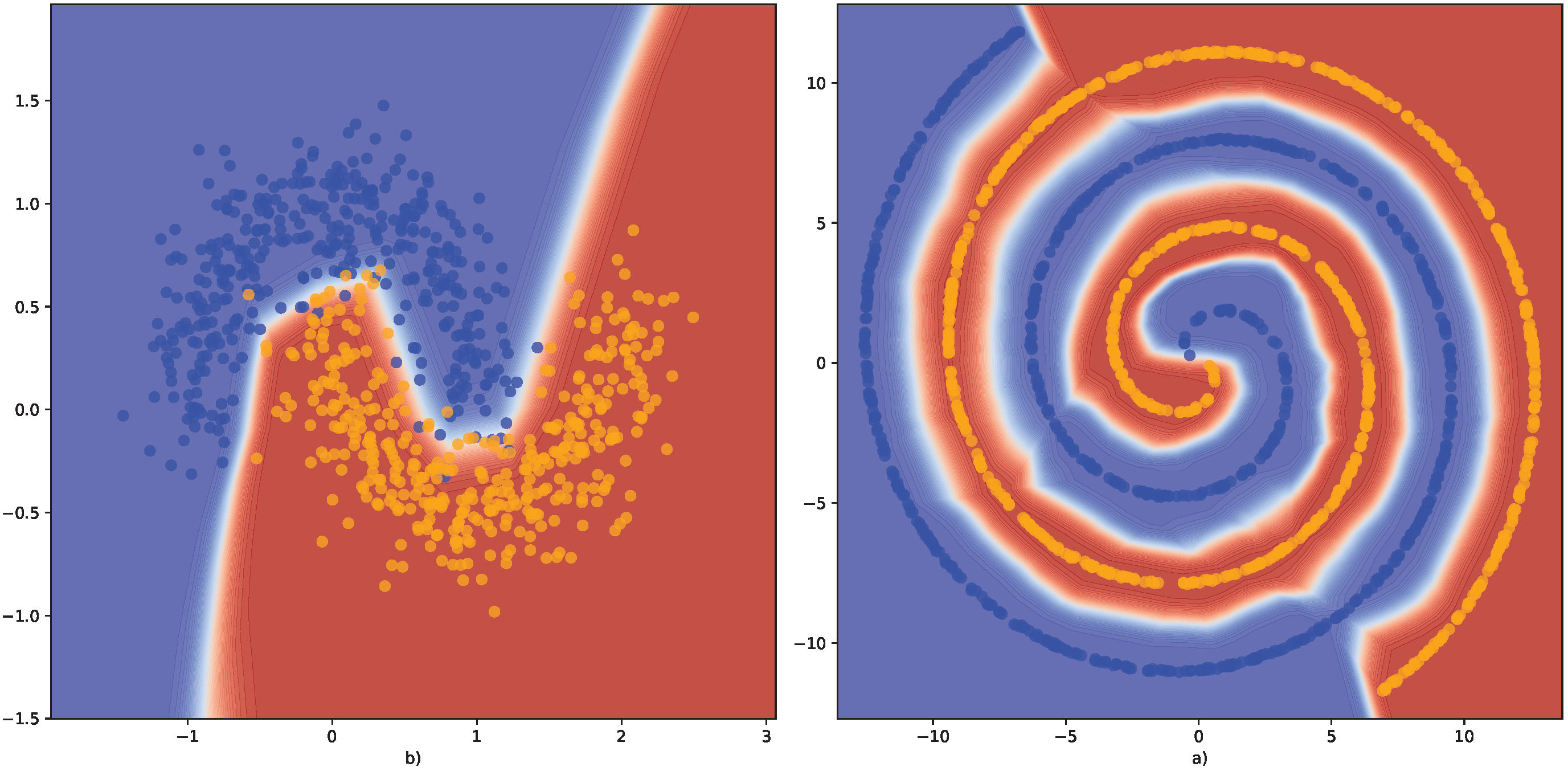}
    \caption{a) Decision boundary for spiral data. b) Decision boundary for moon data.}
    \label{fig:moon&spiral}
\end{figure}

\begin{figure}
    \centering
    \includegraphics[width=0.9\linewidth]{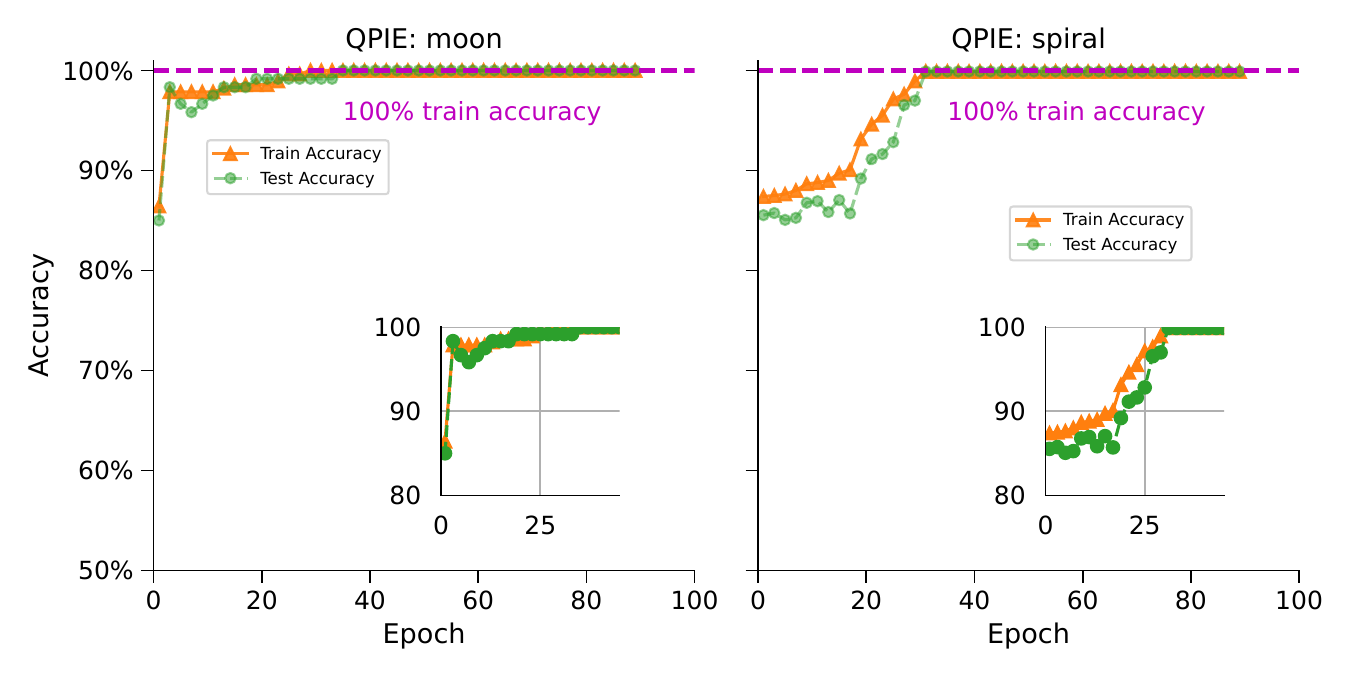}
    \caption{\qp\ benchmark training in 100 epochs with local noisy density matrix CPU 28 qubits quantum simulator.}
    \label{res:qpie}
\end{figure}
\begin{figure}
    \centering   
    \includegraphics[width=0.9\linewidth]{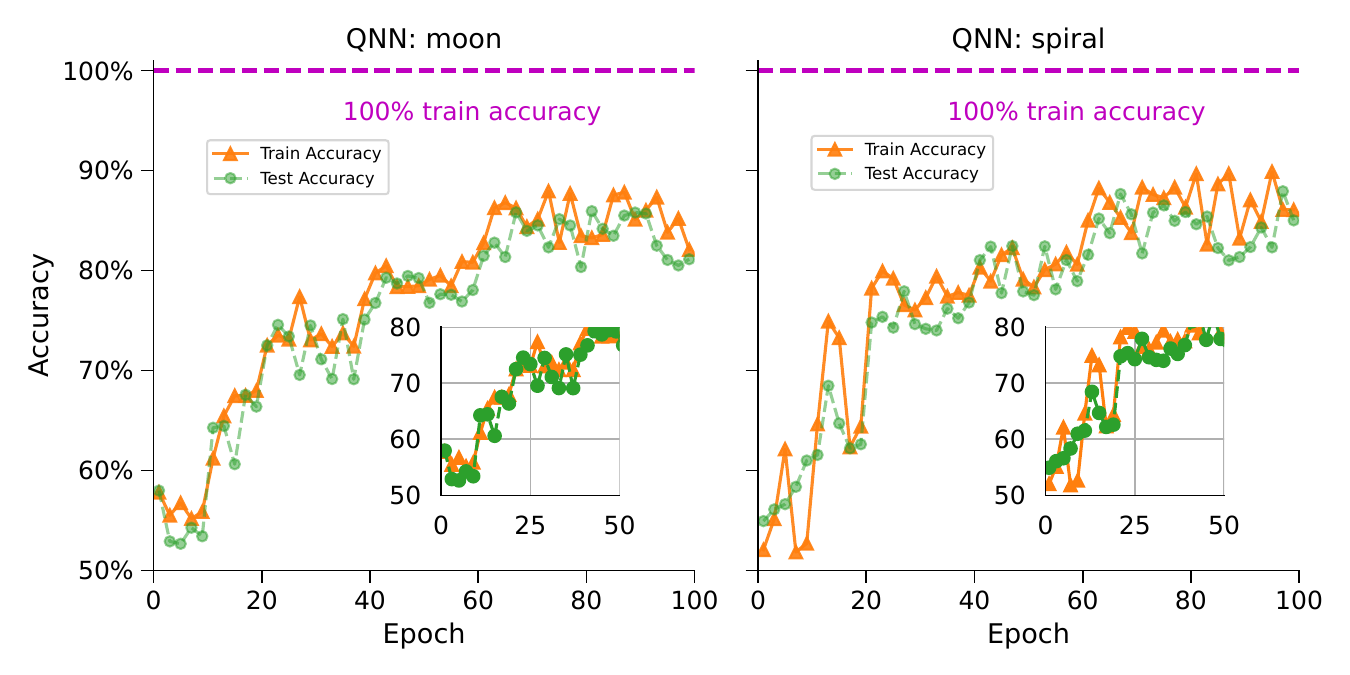}
    \caption{Classical QNN training in 100 epochs with local noisy density matrix CPU 28 qubits quantum simulator (see details settings in \cite{s2}).}
    \label{res:qnn}
\end{figure}

Detailedly, as shown in \cref{fig:moon&spiral}, we showcase over 99\% accuracy in both benchmarking datasets compared to classic NN, as \qp\ can easily discover non-linear patterns in latent space using the commonly used arbitrary rotation gates described in \cref{med:vqc}. Note that the classical component of \qp\ and QNN  reduces classical training time by leveraging fine-tuned weights. In contrast, training a neural network from scratch is required to achieve comparable or higher accuracy, which is more time-consuming.
To further test the model robustness, we introduced a few "outliers" in the moon dataset and observed that the decision boundary was not affected by the human-created noise. However, local simulators have a natural advantage since state vector-based simulators can easily handle up to 28 qubits on a single A100 GPU without noise or overhead issues. In contrast, QPUs suffer from non-eliminable internal noise, causing relatively worse results. 
To account for this, we set our local simulator to use a density matrix CPU backend. This ensures that the final measurements of the VQC not only include the intended bit string outcomes but also incorporate more deviated bit strings (see noise in \cref{fig:noise}). 
We observe that \qp\ demonstrates superior performance in terms of initial accuracy and noise resilience. However, minor oscillations are evident during the first 20 epochs of the training period, as illustrated in \cref{res:qpie} and \cref{res:qnn}. These oscillations are mitigated by the inherent parallel prediction mechanism, which balances certain perturbations, such as phase flips. For example, the application of \(Rx\left(\frac{\pi}{2}\right)\) followed by \(Rx\left(-\frac{\pi}{2}\right)\) can effectively cancel each other out, thereby mitigating quantum errors introduced during execution.



%% file: sections/result_02_series.tex
\subsection*{Time series prediction}

\begin{figure}[htbp]
    \centering
    \includegraphics[width=0.8\linewidth]{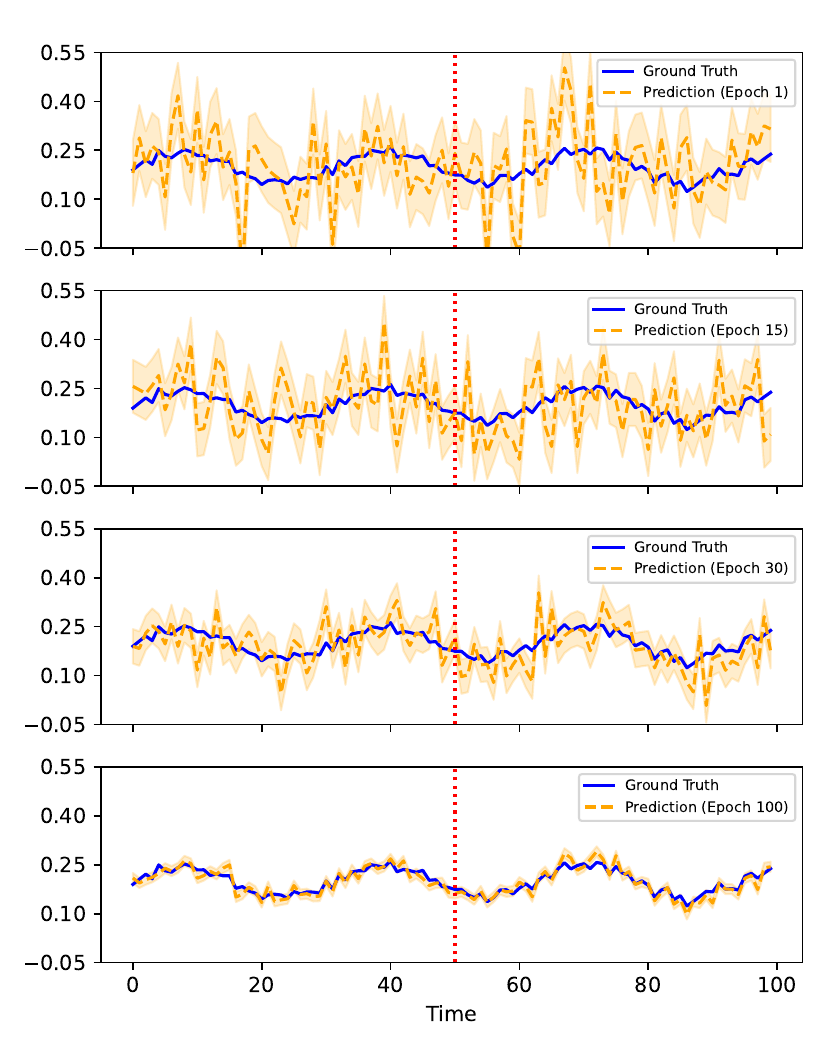}
    \caption{\qp\ result on NARMA5 (100 steps).}
    \label{fig:narma5}
\end{figure}
\begin{figure}[htbp]
    \centering
    \includegraphics[width=0.8\linewidth]{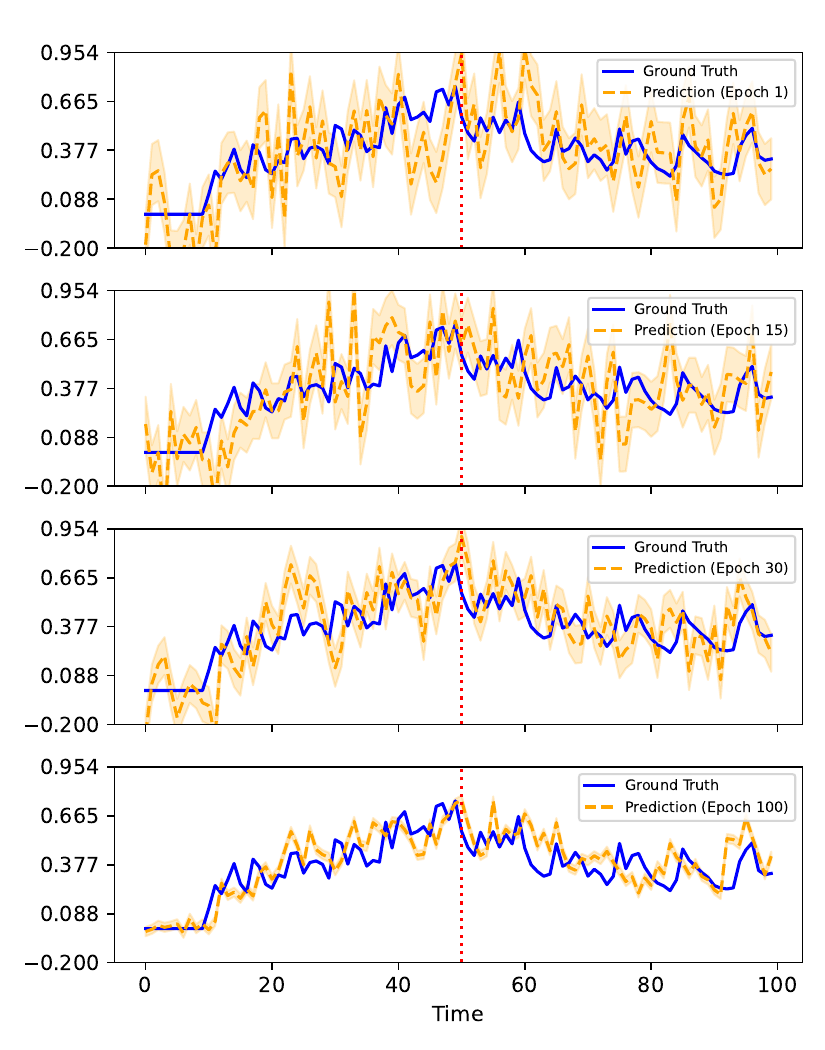}
    \caption{\qp\ result on NARMA10 (100 steps).}
    \label{fig:narma10}
\end{figure}

We examine the \qp\ learning capacity across multiple learning steps utilizing two time-series prediction tasks: NARMA5 and NARMA10 (see \cref{med:dataset} data setup). We employ quantum transfer learning (QTL) (\resnet) combined with \qp\ as described in \cref{tab:results} to capture the NARMA patterns. The advantage of choosing \resnet\ over \resnets\ becomes evident as the quantum circuit depth and number of qubits increase, since \resnet\ contains a relatively smaller proportion of parameters, making it more suitable for transfer learning.
As shown in \cref{fig:narma5} and \cref{fig:narma10}, after 100 training steps, our proposed model successfully predicts the series data with low standard deviation, indicated by the narrow yellow gap in the plots. Notably, for the NARMA10 task, the model demonstrates slightly deferred predictions due to its higher complexity, requiring approximately 83\% more computational effort and 1.5x greater non-linearity compared to NARMA5, as illustrated in \cref{fig:narma10}.
By analyzing the mean squared error (MSE) and standard deviation (SD), we observe that \qp\ achieves a two-magnitude faster learning capacity within only 100 steps in NARMA5 compared to NARMA10.

%% file: sections/result_03_model_capability.tex
\subsection*{Specturm of fisher information}

\begin{figure}
    \centering
    \includegraphics[width=0.98\linewidth]{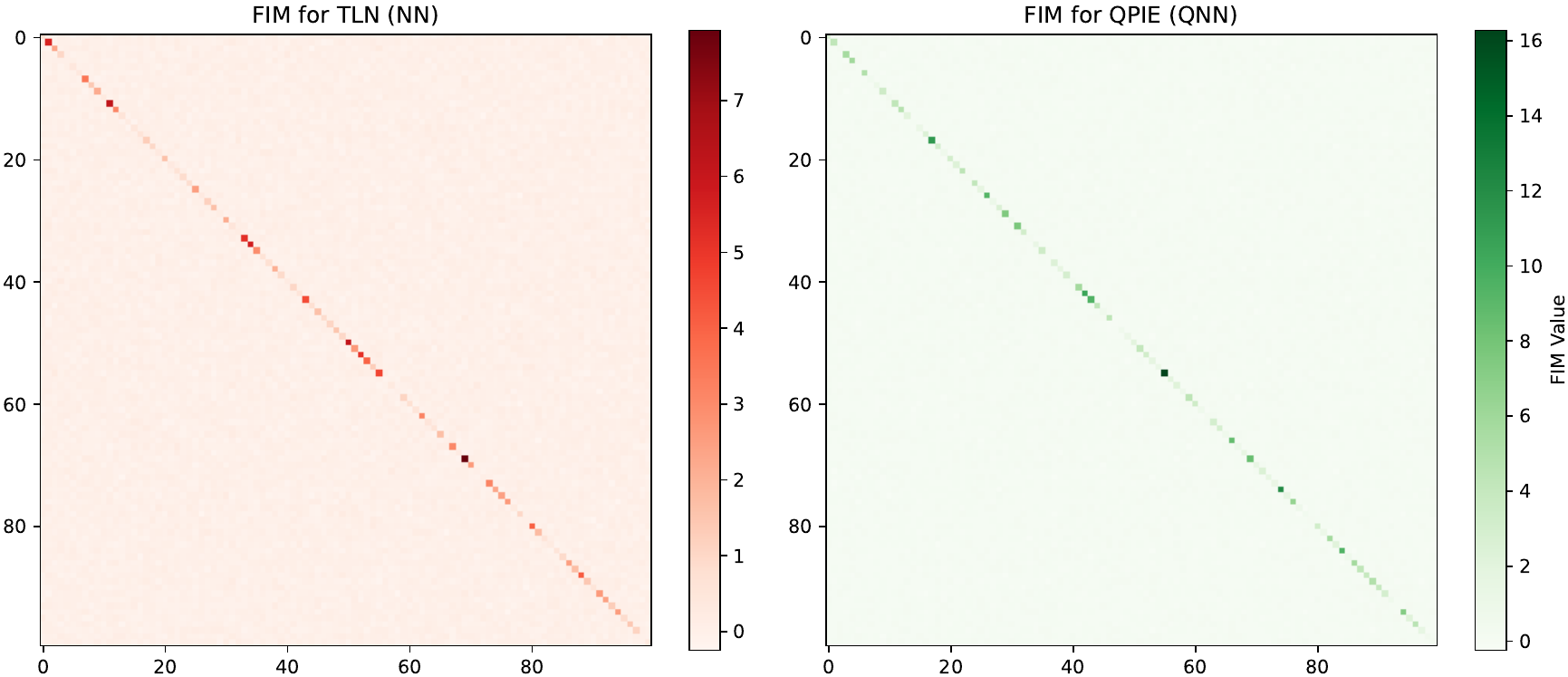}
    \caption{FIM heatmap with 100 parameters input for classical \resnet\ transfer learning network and \qp\ hybrid network.}
    \label{fig:fim}
\end{figure}

\begin{figure}
    \centering
    \includegraphics[width=0.98\linewidth]{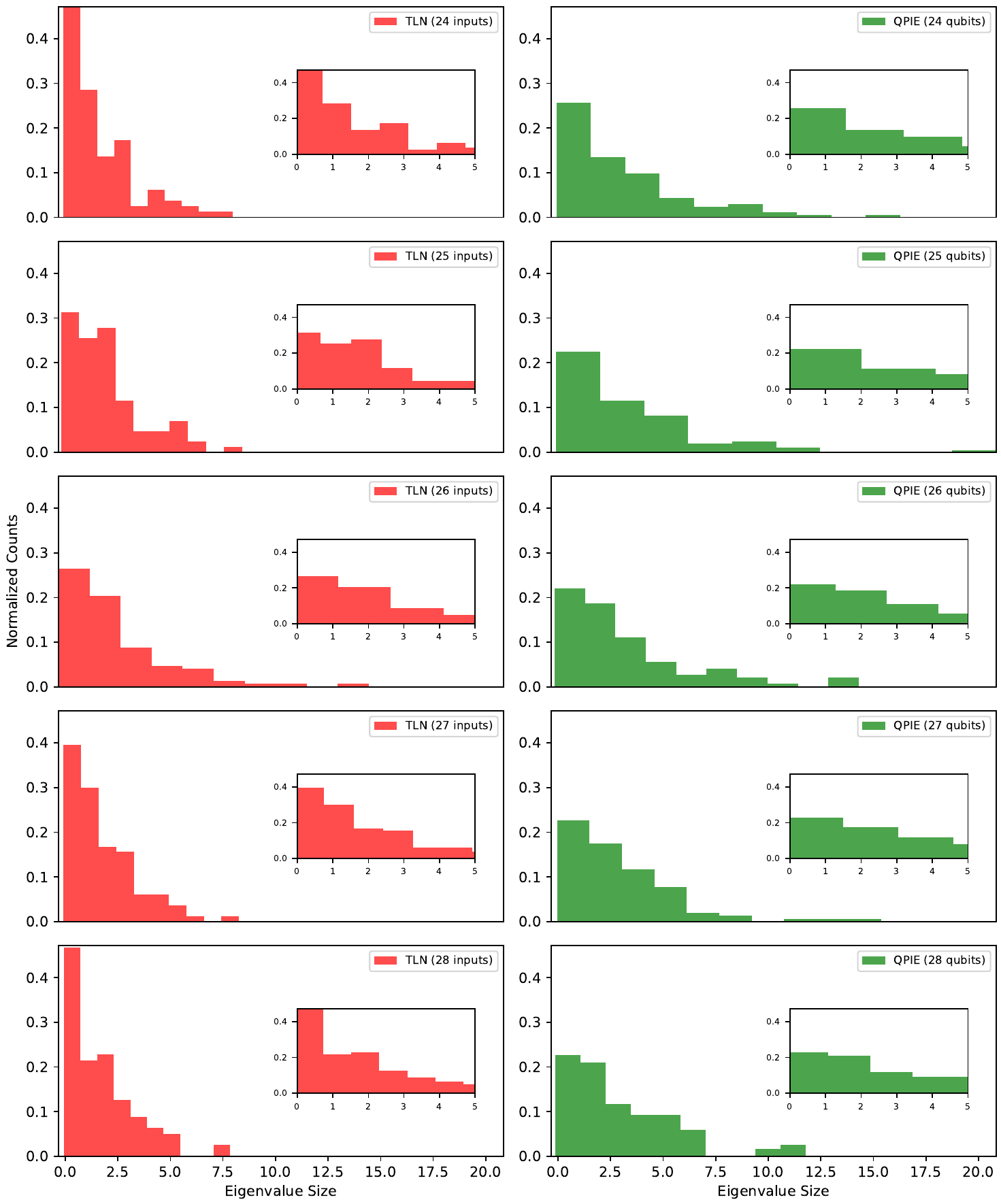}
    \caption{The eigenvalue spectrum of Fisher FIM for classical \resnet\ transfer learning network from 24 to 28 inputs and \qp\ hybrid network with 24 to 28 qubits.}
    \label{fig:eigen}
\end{figure}

The parallel information exchange schedule proposed in this paper leverages state-of-the-art dynamic parameterized quantum circuits (see implementation in \cref{med:vqc}). This enables a larger model capacity and higher eigenvalue sensitivity across the latent high-dimensional space, as shown in \cref{fig:fim}. To evaluate this, we analyze the classical neural inputs from \resnet\ and \qp\ quantum circuits with qubits ranging from 24 to 28, as detailed in \cref{med:dataset}. 
We note that \cref{fig:fim} presents the high-level eigenvalue distribution of \qp\ across the model 100-parameter input space, leveraging parameter re-uploading from QTL. 
We observe that the transfer learning network (TLN) FIM has a smaller eigenvalue range (up to approximately 7), whereas the \qp\ FIM exhibits a much larger range (up to approximately 16) in FIM ranges between the models. This larger range in \qp\ highlights its ability to achieve higher Fisher Information values due to its enhanced latent space representation, enabled by the parallelized multi-layered VQC (see details in \cref{med:vqc}).
These larger eigenvalues suggest that \qp\ parameters are more sensitive to changes, enabling faster learning of patterns, as demonstrated in \cref{fig:eigen}.

Although learning tasks require exponentially more qubits \cite{jerbi2023quantum}, \cref{fig:eigen} shows that \qp\ can detect larger eigenvalues even as the number of qubits increases. With the growth of TLN neuron inputs (tensors) and qubit numbers, the result exhibits a more balanced eigenvalue spectrum in the FIM. This suggests that \qp\ consistently discovers more latent space directions with varying input sizes, as the number of qubits directly impacts the embedding and parameterized rotation gates.
Notably, the evidence of balanced eigenvalue distribution is more pronounced among smaller eigenvalues (ranging from 1 to 5), as highlighted in the insets of \cref{fig:eigen}.
Interestingly, with 25 qubits, \qp\ produces an outlier eigenvalue at 20.0, but this phenomenon disappears when the qubit count increases to 28.
This behavior is attributed to the incorporation of transfer learning in \qp\ architecture, where a flattened quantum Hilbert space often results in near-zero eigenvalues and occasional irrelevant larger eigenvalues \cite{karakida2019universal}.
To mitigate such outliers, we recommend utilizing the maximum number of qubits supported by the available computational resources (CPU, GPU, or QPU) for improved consistency and stability in learning tasks.


%% file: sections/04_methods.tex
\section{Methods}
\label{med}
In this section, we present an overview of \qp\ structure by introducing the general non-sequential hybrid classic quantum model (HCQM) with adaptive optimizer to tackle the hardware limitation and quantum transfer learning (QTL) to speed up the training and improve the performance. Then, we detail the information exchange scheme with parameterized partial entanglement layer (PPEL) and symmetrical angle embedding layer. We also introduce the experimental settings including datasets, hardwares, and platforms.

\begin{figure}
    \centering
    \includegraphics[width=0.98\linewidth]{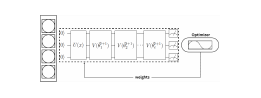}
    \caption{General QNN architecture.}
    \label{fig:hcqm}
\end{figure}

\subsection*{Non sequential hybrid classic quantum model}
\label{med:hcqm}
Based on experiments using a density matrix noisy simulator \cite{Kim2023cudaq, pennylane}, we modify the connecting layer utilizing non-sequential HCQM adapted from sequential QNN structure \cite{s2} to increase the learning task accuracy by distributing each quantum measured results utilizing each quantum node or cuda quantum kernel \cite{Kim2023cudaq} instead of calculating the results one-by-one to reduce the potential outliers gained from training steps, which leads to the robust resistance for noisy simulator provided by \amazon\ \sv.\ 

\cref{fig:hcqm} illustrates the architecture in which classical neurons (inputs) are interconnected with multiple quantum gates. Notably, each quantum gate is parameterized by classical numerical values.
We use a classical optimizer to minimize the training cost within the hybrid network, where quantum gradients are calculated using the parameter-shift rule (PSR) \cite{shift} and adjoint differentiation \cite{adjoint}. Specifically, we define two gradient computation schemes: local quantum simulators use adjoint differentiation, while cloud QPUs and CPUs rely on PSR.
Adjoint differentiation computes gradients by leveraging reverse-mode differentiation, which scales linearly with the iteration training as it requires only a forward and backward gradient calculation of the circuit.
In our real QPU experiment, the parameter-shift rule calculates gradients by running the circuit twice per parameter with shifted values (\( \theta + s \) and \( \theta - s \)). We set \(s = \frac{\pi}{2}\) because it provides precise derivatives for the trigonometric dependence. The gradients calculation is shown by \cref{eq: grad}
\begin{equation}
    \frac{\partial \langle M \rangle}{\partial \theta} = \frac{1}{2} \left( \langle M \rangle_{\theta + \pi/2} - \langle M \rangle_{\theta - \pi/2} \right).
    \label{eq: grad}
\end{equation}
But, for our learning task, with a maximum of 784 parameters (from 28x28 pixels), the linear scaling of adjoint differentiation accelerates training on local simulators.
\[
\begin{array}{c}
\text{Input}\\
\Big\downarrow \\[4pt]
\begin{array}{ccc}
\mathcal{C}_1 \to \mathcal{Q}_1 & \quad \mathcal{C}_2 \to \mathcal{Q}_2 & \quad \mathcal{C}_3 \to \mathcal{Q}_3
\end{array} \\[8pt]
\Big\downarrow \\[4pt]
\text{Output}
\end{array}
\]
\vskip -0.15in
\begin{figure}
    \centering
    \includegraphics[width=0.98\linewidth]{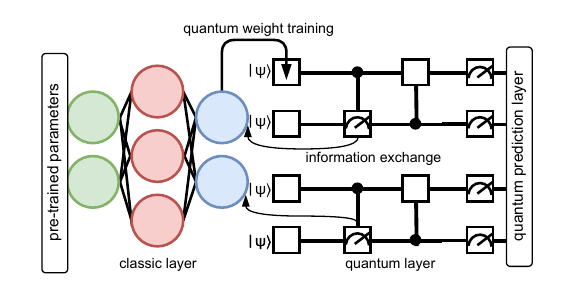}
    \caption{A \qp\ hybrid network with quantum transfer learning, where the classical neuron and quantum layer can be trained in parallel using GPUs and QPUs.}
    \label{fig:qpie}
\end{figure}
\subsection*{Quantum parallel exchange information non-sequential network}
\label{med:qpie}
%
In multiple non-sequential VQCs indicated by two quantum circuits in \cref{fig:qpie}, we represent the classical network output as the "Input" label above three parallel VQCs, denoted as (\(\mathcal{Q}_1, \mathcal{Q}_2, \mathcal{Q}_3\)), which merge into a single "Output" computed by the quantum prediction layer. \qp\ enables dynamic mid-circuit measurement, which collapses the quantum state into a classical state, allowing the quantum layer to exchange gate rotation parameters with the classical layer based on a predefined measurement threshold.
Specifically, we define a one-hot matrix to represent the rotation type pool, which determines the applicable quantum rotation gates (\(R_x, R_y, R_z\)).
\[
\text{Rotation Type Pool: }
\begin{bmatrix}
1 & 0 & 0 \\
0 & 1 & 0 \\
0 & 0 & 1
\end{bmatrix}.
\]
The rotation types are dynamically selected based on a measurement value (\(\text{meas}\)) using the conditional function defined by \cref{eq: rt}
\begin{equation}
    g(\Theta) =
\begin{cases} 
R_x(\Theta), & \text{if } \text{meas} < \tau_1, \\
R_y(\Theta), & \text{if } \tau_1 \leq \text{meas} < \tau_2, \\
R_z(\Theta), & \text{if } \text{meas} \geq \tau_2.
\end{cases}
\label{eq: rt}
\end{equation}
We define \(\tau_i = \text{G}(\text{meas}, q_i)\), where \(q_i\) denotes the quantum state corresponding to each unitary rotation indexed by \(i\), derived from the modulo-three of the total number of gates and \(\theta\)s are the embedded rotation angles transformed by the classic inputs.

\subsection*{Variational quantum circuit}
\label{med:vqc}
\begin{figure*}[htbp]
    \centering
    \includegraphics[width=0.98\linewidth]{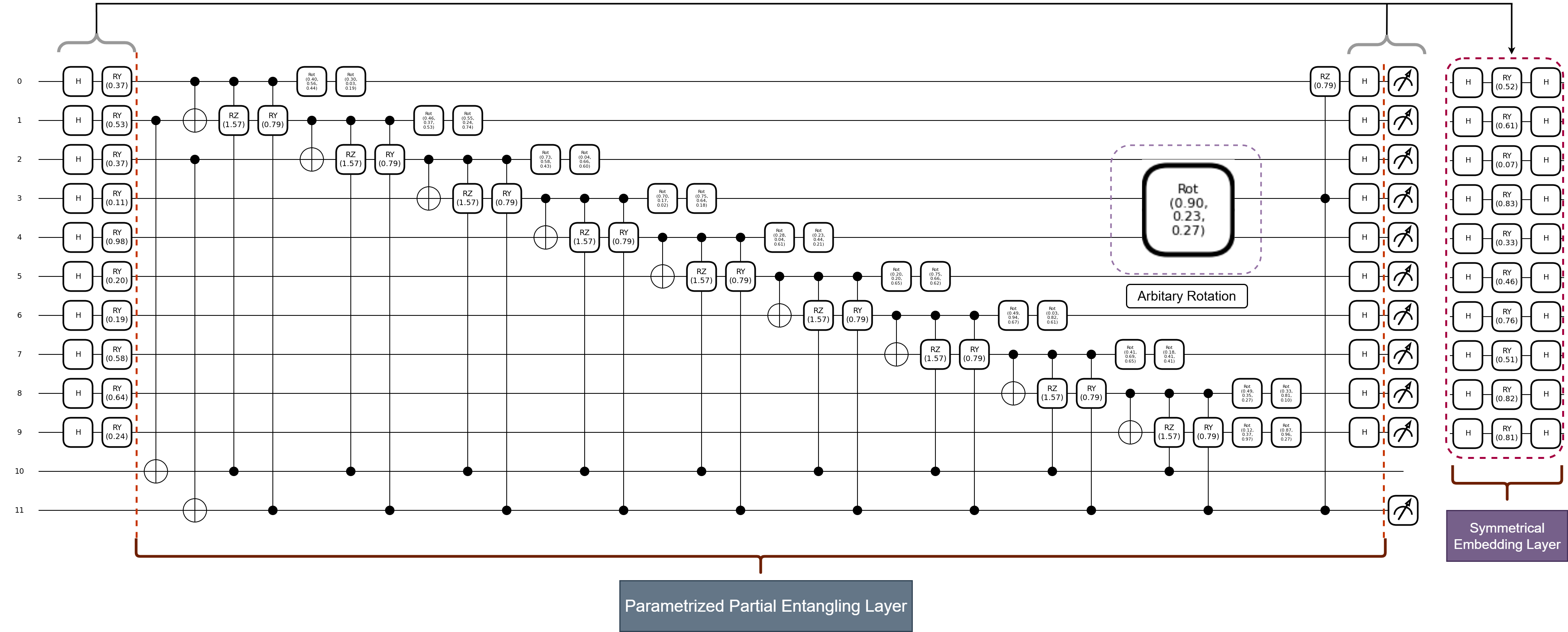}
    \caption{We present the case of 10 learning qubits with 2 ancilla qubits VQC.\ The symmetrical embedding layer (SEL) is composed by the red dashed box involving two Hadamard layers that creates unbiased parameters differentiation and one Y angle embedding layer that provides complex number feature extraction. In the parameterized partial entanglement layer (PPEL), we choose arbitrary gates (three-axis rotation gates) corresponding to the dynamic rotation pool to save the circuit depth and controlled rotation gates (non-clifford gates) to change each qubit angle based on previous measurement. We select two ancillary qubits and measure the last qubit to decide whether we exchange the quantum information with classic layer, where 0 represents off and 1 indicates on.}
    \label{fig:VQC}
\end{figure*}

In \cref{fig:VQC}, we showcase the VQC structure that leverages 10 data qubits and 2 ancilla qubits to efficiently control the data exchanging, with mid-circuit measurements enabling dynamic gradient feedback for hybrid optimization.
The PPEL enhances expressiveness and gradient differentiation, as non-Clifford gates can be efficiently transpiled into universal unitary operations on QPUs~\cite{cross2016scalable}.
We normalize the flattened classical features \(x_1, x_2, x_3\) to fit the parameter ranges of the \(R_3\) gate as follows: \(\theta = \text{normalize}(x_1, [0, \pi])\), \(\phi = \text{normalize}(x_2, [0, 2\pi])\), and \(\lambda = \text{normalize}(x_3, [0, 2\pi])\).
These normalized parameters control the quantum rotation in the \(R_3\) gate, defined as \cref{eq: r3} marked by the purple dashed box.
\begin{equation}
   R_3(\theta, \phi, \lambda) = \begin{bmatrix} \cos\left(\frac{\theta}{2}\right) & -e^{i\lambda} \sin\left(\frac{\theta}{2}\right) \\ e^{i\phi} \sin\left(\frac{\theta}{2}\right) & e^{i(\phi+\lambda)} \cos\left(\frac{\theta}{2}\right) \end{bmatrix}. 
   \label{eq: r3}
\end{equation}

This leads the dynamic mapping of classical features to quantum states.
Furthermore, the PPEL selectively entangles qubits (without the entanglement of ancilla qubits) to balance noise mitigation and avoid the decoherence issues of full entanglement as introduced by final qubit measurement to control the switch of exchange information. Lastly, we separate the symmetrical embedding layers—one placed at the beginning and the other at the end—to ensure uniform and adaptive data encoding because the non-consecutive avoids placing consecutive Hadamard layers together, which would otherwise reduce to the identity operation and contribute no meaningful transformation.
\subsection*{Experiment}
\label{med:exp}
We choose the fully connected neural network settings for binary learning task as indicated by \cite{nn}. 
The classical layer consists of five fully connected layers, each employing SeLU activations \cite{selu}. Dropout is applied after the second and fifth layers to mitigate the overfitting problem.
In our cases, we utilize \resnet\ and \resnets\ as the pre-trained models before connecting to the quantum layer \cite{mari2020transfer}. For \resnet\ and \resnets\ setup, we maximumly implement 28 qubits VQC with encrypting \(2^{28}\) different states that are equivalent to \(\sim 45\% \) pre-trained parameters. Therefore, we initially freeze the first two stages learnt parameters and dynamically connect the unlocked parameters with our VQCs, where the weights are fed into variational quantum gates, utilizing ancilla qubits proportional to the logarithm of the number of final predictions, defined as \( n_q = \lceil \log_2(N_c) \rceil \). Consequently, this ensures that the prediction task scales logarithmically with the number of classes, making the system computationally efficient.

We then use the \pen\ and \cq\ quantum simulation framework for quantum node creation because \pen\ provides flexible quantum parameters differentiation and \cq\ enables cuquantum backend~\footnote{https://docs.nvidia.com/cuda/cuquantum/latest/index.html} GPU speed-up. We run hybrid job experiments on \amazon\ Bracket \cite{braket} that enable different platform selections such as \sv\ and \aria.\ The local simulator experiment with the Toreador 11 system with AMD CPU and four NVIDIA A100 GPUs distributed within one node; each GPU has 11.8 GB of GPU memory with total of 47.2 GB.

\begin{figure}
    \centering
    \includegraphics[width=1.0\linewidth]{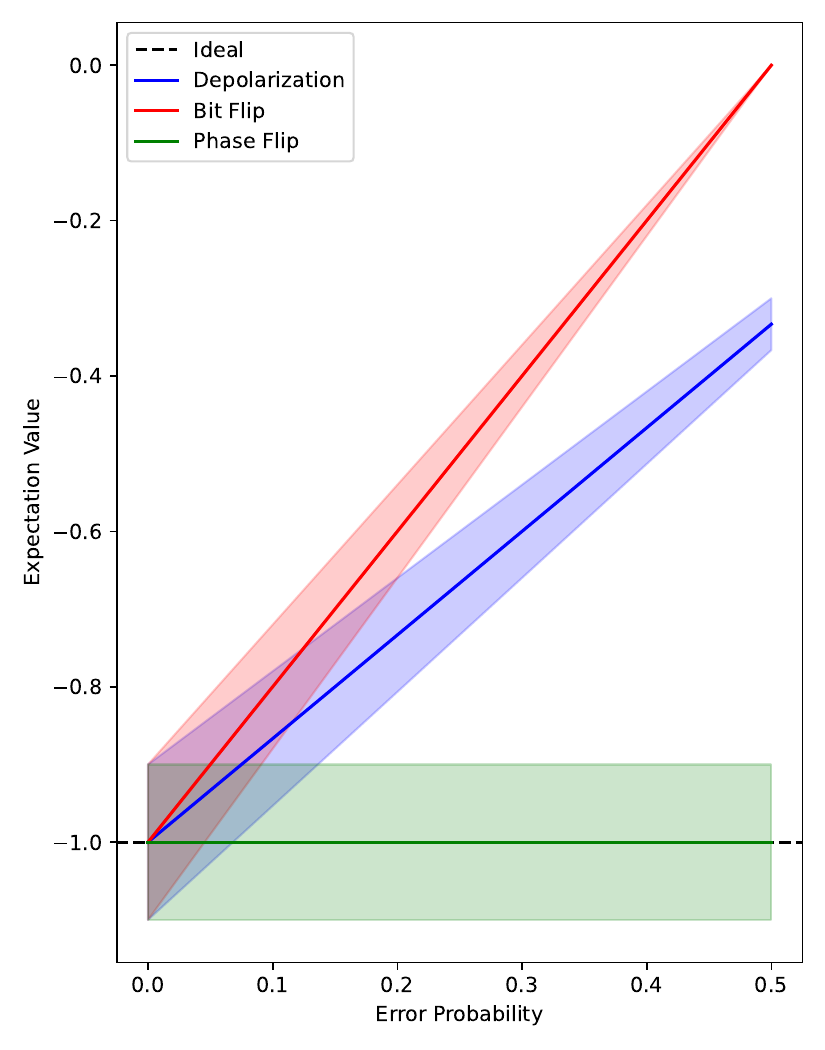}
    \caption{The expectation value of measured bit strings corresponding to the depolarization, bit flip, and phase flip error probability.}
    \label{fig:noise}
 
\end{figure}

In a SV1 state vector quantum simulator, we modify the number of shots to \(2^n\), where \(n\) is the number of qubits, ensuring that all possible measurement outcomes are sampled exactly once for a deterministic circuit with statistical noise (see \cref{fig:noise}) proving the mid-circuit interactive model has higher fidelity \cite{mid}. 
We note this leads to a feedback mechanism where the mid-circuit quantum measurements dynamically influence the classical network so that unfrozen parameters can refine the deeper layers.

\subsection*{Datasets}
\label{med:dataset}
We first select two standard non-linear benchmark datasets (moon and spiral) and three image datasets (hymenoptera, brain tumors, and mnist) requiring sophisticated non-linear pattern learning.
We then choose NARMA5 and NARMA10 as series prediction tasks, where NARMA10 exhibits significantly higher complexity compared to NARMA5. Specifically, for both datasets, we add exponentially decaying noise as shown by \cref{eq: noise}.
\begin{equation}
    \text{Prediction}_{\text{epoch}}(t) = y(t) + \eta \epsilon(t), \quad \eta = \alpha e^{-\text{epoch}/50},
    \label{eq: noise}
\end{equation}
where \(\epsilon(t) \sim \mathcal{N}(0, 1)\), and \(\alpha = 0.1\) for NARMA5 and \(\alpha = 0.2\) for NARMA10. The standard deviation of the noise is denoted as \(\sigma = \text{std}(\eta \epsilon(t))\).
We use the FIM to evaluate parameter sensitivity for the TLN (classical NN) and QPIE (quantum NN) models. For both models, the eigenvalue density \(P(\lambda)\) is given by \cref{eq: fim}.
\begin{equation}
    P(\lambda) = \frac{\text{Count}(\lambda)}{\text{Total Count}}, \quad \lambda > 0,
    \label{eq: fim}
\end{equation}
where \(\lambda\) denotes the eigenvalues of the FIM.



%% file: sections/06_related_work.tex
\section{Related Work}
Hybrid Quantum Computing (HQC) seeks to discover optimal quantum circuits for efficient feature extraction \cite{b11}, uncover novel representations for the hybrid shallow quantum circuits approach \cite{hybird1}, and address complex tasks such as high dimensional data analysis \cite{qsvm}, quantum pixel reconstruction \cite{amankwah2022quantum}, and quantum error mitigation \cite{b8}. Furthermore, quantum RNN-based methods \cite{bausch2020recurrent} enable more accurate and larger capacity in time-series and real-world modeling tasks \cite{transfer1}, utilizing algorithms like Quantum Phase Estimation (QPE), Quantum Fourier Transform (QFT), and Quantum Embedding (QE) \cite{dorner2009optimal, qe}. The proposed \qp\ framework aims to leverage efficient hybrid classic quantum model with transfer learning while enhancing the training efficiency and noise robustness, contributing to solve the learning tasks with current gate-based mathematical representation of quantum circuit learning in Hilbert space \cite{schuld2019quantum}.

%% file: sections/03_discussion.tex
\section{Discussion}
In this paper, we develop a \qp\ hybrid network with transfer learning, designed to optimize the performance of quantum circuit learning (QCL) within the constraints of NISQ devices, outperforming conventional QNNs. 
In summary, this study presents three main contributions:
1) We demonstrate that \qp, with pre-trained weights, provides an efficient framework for learning tasks.
2) We propose a dynamic information exchange hybrid network that is migratable to current QNN-based models. Furthermore, we prove that this scheme mitigates training fluctuation issues under noisy quantum simulators by leveraging parameterized quantum circuits, quantum embedding, and parallel training.
3) Our experimental results validate \qp\ on the trapped ion QPU, showcasing the versatility of our approach across different hardware. While promising results are achieved on real QPUs, we acknowledge that these outcomes benefit from curated quantum error suppression techniques available on current platforms \cite{ball2021real}.

%% file: sections/acknowledgements.tex
\section*{Acknowledgements}
We appreciate Xanadu for sponsoring us with AWS Bracket credits for accessing the real hardware resources and acknowledging using Pennylane and AWS local SV1 state vector simulator, and Aria1 backbones.
The authors acknowledge the High Performance Computing Center (HPCC) at Texas Tech University for providing computational resources that have contributed to the research results reported within this paper. URL: http://www.hpcc.ttu.edu
